\documentclass[aps,prl,twocolumn,groupedaddress]{revtex4-1}

\usepackage[pdftex]{graphicx}
\usepackage{epstopdf}
\usepackage{amsmath}
\usepackage{color}
\usepackage{siunitx}
\usepackage{textcomp}
\usepackage[T1]{fontenc}

\begin{document}


\title{Superconductivity in infinite-layer CaCuO$_{2}$-brownmillerite Ca$_{2}$Fe$_{2}$O$_{5}$ superlattices}


\author{Ai Ikeda, Yoshiharu Krockenberger, Yoshitaka Taniyasu, and Hideki Yamamoto}
\affiliation{NTT Basic Research Laboratories, NTT Corporation, 3-1 Morinosato-Wakamiya, Atsugi, Kanagawa 243-0198, Japan}


\date{\today}

\begin{abstract}
High-temperature cuprate superconductors have naturally a superlattice structure. Infinite-layer CaCuO$_{2}$ is the common ingredient of cuprates with superconducting transition temperatures above 100\,K. However, infinite-layer CaCuO$_{2}$ by itself does not superconduct. Here we show that superconductivity emerges in artificial superlattices built from infinite-layer CaCuO$_{2}$ and brownmillerite Ca$_{2}$Fe$_{2}$O$_{5}$ grown by molecular beam epitaxy. X-ray diffraction and electron microscopy characterizations showed that the crystal quality of the infinite-layer CaCuO$_{2}$ in the superlattices significantly improved compared to bare thin-films of CaCuO$_{2}$. We found that the induction of superconductivity in [(CaCuO$_{2}$)$_{n}$(Ca$_{2}$Fe$_{2}$O$_{5}$)$_{m}$]$^{N}$ superlattices is also subject to the oxidizing environment used during the cool-down procedure and therefore to a minimization of oxygen vacancies within the CuO$_{2}$ planes. The inserted Ca$_{2}$Fe$_{2}$O$_{5}$ layers buffer charge imbalances triggered by point defect formation during growth, minimizing cationic defects in the infinite-layer CaCuO$_{2}$ layers thus stabilizing monolithic infinite-layer CaCuO$_{2}$ slabs; embedding CaCuO$_{2}$ within a superlattice enables extended two-dimensional CuO$_{2}$ planes and therefore superconductivity while Ca$_{2}$Fe$_{2}$O$_{5}$ serves similar to the charge-reservoir layers in the cuprate superconductors synthesized from incongruent melts.

\end{abstract}


\maketitle
\section{INTRODUCTION}
The layered structure is one of the key features found in high-temperature superconductors, such as cuprates and iron-pnictides. Their structures can be simply regarded as alternate stacking of superconducting layers, $e.g.$, (CuO$_{2}$)$^{2-}$ or (FeAs)$^{-}$, and other building blocks, so-called charge reservoir layers (CRLs), adapted to fulfill the charge neutrality criteria. Mimicking these superconducting layered materials, heteroepitaxy has a great potential to design and create novel superconductors by means of artificial superlattices. In oxide systems, SrTiO$_{3}$/LaAlO$_{3}$ superlattices/heterostructures is one example but its superconducting two-dimensional electron system is confined at the interfaces between SrTiO$_{3}$ and LaAlO$_{3}$ \cite{Ohtomo2004, Reyren1196}. Nonetheless, using infinite-layer (IL) CaCuO$_{2}$ instead of SrTiO$_{3}$ sandwiched by another oxide might be promising to design new superconductors.

\begin{figure}
\includegraphics[width=8.6cm]{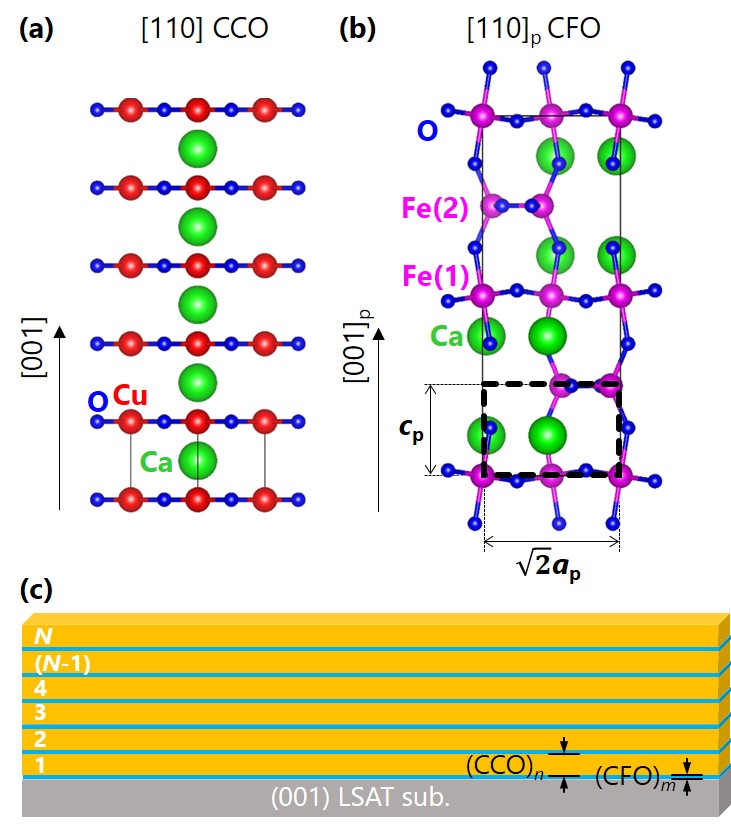}%
\caption{Crystal structures of (a) IL-CaCuO$_{2}$ (CCO), (b) brownmillerite Ca$_{2}$Fe$_{2}$O$_{5}$ (CFO), and (c) a scheme of [(IL-CCO)$_{n}$/(CFO)$_{m}$]$^{N}$ superlattices. In (b), the perovskite subcell axis with $a_{\text{p}}$ and $c_{\text{p}}$ are shown by dashed lines. The perovskite subcell axis is used for the determination of $m$. \label{fig1}}
\end{figure}

IL-cuprates ACuO$_{2}$ (A = Ca, Sr), where simple cations A$^{2+}$ separate adjacent square-planar CuO$_{2}$ planes, are the only cuprates without CRLs [Fig.\ref{fig1}(a)]. In multilayer cuprates with more than three CuO$_{2}$ planes in a unit cell, $e.g.$, Bi$_{2}$Sr$_{2}$Ca$_{2}$Cu$_{3}$O$_{10+\delta}$, HgBa$_{2}$Ca$_{2}$Cu$_{3}$O$_{8+\delta}$, and TlBa$_{2}$Ca$_{2}$Cu$_{3}$O$_{8+\delta}$, the inner square-planar CuO$_{2}$ plane sandwiched between Ca$^{2+}$ planes commonly forms the IL structure whereas CRLs are composed of different oxide layers. $T_{\text{c}}$ of the multilayer cuprates by far exceed 100\,K and such high $T_{\text{c}}$-superconductivity is immanent in the CuO$_{2}$ planes in the IL-CaCuO$_{2}$ layers. IL-CaCuO$_{2}$ itself, however, is known to be an antiferromagnetic charge-transfer insulator. In our earlier studies, we have shown that the absence of superconductivity in IL-CaCuO$_{2}$ and electron doped IL-Ca$_{1-x}$R$_{x}$CuO$_{2}$ (R: trivalent rare-earth) is associated with the presence of anti-site boundaries along the [100] or [010] directions, where CuO$_{2}$ planes are disconnected \cite{PhysRevMaterials.3.064803}. The formation of such defects most likely originates from the polar character of Ca-based IL-cuprates, which appears to be an inherent problem in this system. In contrast, superconductivity was found in superlattices/hetrostructures of IL-CaCuO$_{2}$ and perovskite SrTiO$_{3}$, two insulating materials \cite{PhysRevB.86.134524,doi:10.1063/1.4768680}. Although the mechanism of superconductivity observed in (IL-CaCuO$_{2}$)$_{n}$/(SrTiO$_{3}$)$_{m}$ superlattices is unclear, one can expect that exchanging SrTiO$_{3}$ by other oxides may meet the requirements to induce superconductivity in IL-CaCuO$_{2}$ and furthermore pushing up the limit of $T_{\text{c}}$ by designing the superlattices.

Synthesizing the desired superlattice structure with IL-CaCuO$_{2}$ is challenging because of thermodynamic instability of the IL phase. High-pressure synthesis is required to synthesize IL-ACuO$_{2}$ in bulk form.  Nonetheless, we have shown earlier that molecular beam epitaxy (MBE) is a powerful method to synthesize single crystalline thin films of the IL phase \cite{APEX5.043101}. The synthesis of the IL-cuprates in a thin-film form requires the use of well-lattice-matched substrates and strong oxidizing agents such as atomic oxygen/ozone \cite{JM9950501879, doi:10.1063/1.4767117, JAP.124.2018.073905}. Under consideration of these constraints of the growth of IL-CaCuO$_{2}$, we targeted brownmillerite Ca$_{2}$Fe$_{2}$O$_{5}$ for the following reasons: (a) small lattice mismatching between Ca$_{2}$Fe$_{2}$O$_{5}$ and IL-CaCuO$_{2}$ ($\sim$1.1$\%$), (b) thermodynamic compatibility between Ca$_{2}$Fe$_{2}$O$_{5}$ and IL-CaCuO$_{2}$. In addition, Ca$_{2}$Fe$_{2}$O$_{5}$ is a well-known insulator with a N\'{e}el temperature of ~720 K \cite{doi:10.1143/JPSJ.24.446}. This allows us to eliminate electronic contributions from Ca$_{2}$Fe$_{2}$O$_{5}$ layers to the overall conductivity of the synthesized superlattices. There have been reports where perovskites were paired with brownmillerites in superlattices \cite{Matsumoto2011, Jo2020}. However, the combination of IL and brownmillerite oxides remains unexplored.


In this work, we synthesized [(IL-CaCuO$_{2}$)$_{n}$/(Ca$_{2}$Fe$_{2}$O$_{5}$)$_{m}$]$^{N}$ superlattices by MBE to search for superconductivity in this system. Using the state-of-the-art electron microscopy measurements, we confirmed cation as well as anion arrangements in IL-CaCuO$_{2}$ and brownmillerite Ca$_{2}$Fe$_{2}$O$_{5}$ layers within the superlattices at the atomic level. Resistivity and magnetization measurements revealed bulk nature of superconductivity in the [(IL-CaCuO$_{2}$)$_{n}$/(Ca$_{2}$Fe$_{2}$O$_{5}$)$_{m}$]$^{N}$ superlattices. The superlattice approach presented here allowed us to stabilize IL-CaCuO$_{2}$ without significant structural distortion and utilize it for the synthesis of designer superconducting cuprates.

\section{EXPERIMENTAL DETAILS}
Artificial superlattices of [(IL-CaCuO$_{2}$)$_{n}$/(Ca$_{2}$Fe$_{2}$O$_{5}$)$_{m}$]$^{N}$ were synthesized on (001) (LaAlO$_{3}$)$_{0.3}$(SrAl$_{0.5}$Ta$_{0.5}$O$_{3}$)$_{0.7}$ (LSAT) substrates by MBE. Our custum designed MBE system was equipped with 10 elemental sources and six electron guns with source shutters. Co-evaporation of atomic beam fluxes of Ca, Cu, and Fe was controlled by electron impact emission spectroscopy (EIES) \cite{YAMAMOTO2013184}. The source shutter timing was programmed in conjunction with real-time monitoring of flux rates. The typical growth rate for the synthesis of superlattices was approximately 1\,\AA/s. The surface structure of each layer was monitored $in\,situ$ by reflection high energy electron diffraction (RHEED). A custom designed atomic oxygen source operated at 13.56\,MHz (radio frequency RF) was used to provide a strong oxidizing environment. The RF power of the atomic oxygen source and oxygen flow rate used are 300\,W and 1.5\,sccm (an oxygen background pressure of 4$\times$10$^{-6}$\,Torr), respectively. The substrate temperature ($T_{\text{s}}$) was measured by a radiation pyrometer (Japan Sensor). $T_{\text{s}}$ of 580-590\textdegree C was used to synthesize [(IL-CaCuO$_{2}$)$_{n}$/(Ca$_{2}$Fe$_{2}$O$_{5}$)$_{m}$]$^{N}$ superlattices. We first grew a Ca$_{2}$Fe$_{2}$O$_{5}$ layer of $m$ unit cells followed by an IL-CaCuO$_{2}$ layer of $n$ unit cells and repeated this $N$ times [Fig. \ref{fig1}(c)]. The $N$ value varied between 6 and 15 in order to obtain a total film thickness of 400-800\,\AA . The thickness of IL-CaCuO$_{2}$ layers ($t_{\text{CCO}}$) was set in the range of 40 to 55\,\AA{ }corresponding to $n$ from 13 to 17 unit cells, while that of Ca$_{2}$Fe$_{2}$O$_{5}$ layers ($t_{\text{CFO}}$) varied between 7 and 18\,\AA{ }corresponding to $m$ between 2 and 5 unit cells. Note that we use the prerovskite subcell axis ($c_{\text{p}}$) to determine $m$ for Ca$_{2}$Fe$_{2}$O$_{5}$ [see Fig.\ref{fig1}(b)]. After the growth, the superlattices were cooled down to $T_{\text{ox}}$ $\leq$ 140\textdegree C under flowing atomic oxygen with the RF power of 300\,W and oxygen flow rate of 0.8\,sccm.

The structural characterization was carried out by high-resolution X-ray diffraction (XRD) on Bruker D8 4-circle diffractometer. In addition, aberration-corrected scanning transmission electron microscopy (STEM) combined with electron energy loss spectroscopy (EELS) measurements was performed on a JEOL ARM-200F electron microscope equipped with an electron energy loss spectrometer (Gatan). Specimens were prepared using ion beam milling for in-plane and cross-sectional STEM measurements. High-angle annular dark field (HAADF) images were collected with a rate of 0.5 - 1.0\,s/image, and integrated to minimize image shifts during acquisition \cite{KIMOTO2010778, JAP123.064102.2018}.

Resistivity ($\rho$) measurements were performed using a standard four-probe method with silver electrodes. Temperature dependencies of $\rho$($T$) were measured between 300 and 4\,K. Magneto-resistivity between 200 and 1.8\,K was measured in a Quantum Design Dynacool physical property measurement system under magnetic fields up to 14\,T applied perpendicular to the film surface. Au wires (50 \si{\micro}m) were bond to the silver electrodes on the superlattice films using indium. Magnetization measurements were performed using a Quantum Design MPMS3 SQUID-VSM magnetometer. An external magnetic field of 5\,Oe was applied parallel to the film surface to minimize extrinsic effects on magnetization.

\begin{figure*}
\includegraphics[width=17.2cm]{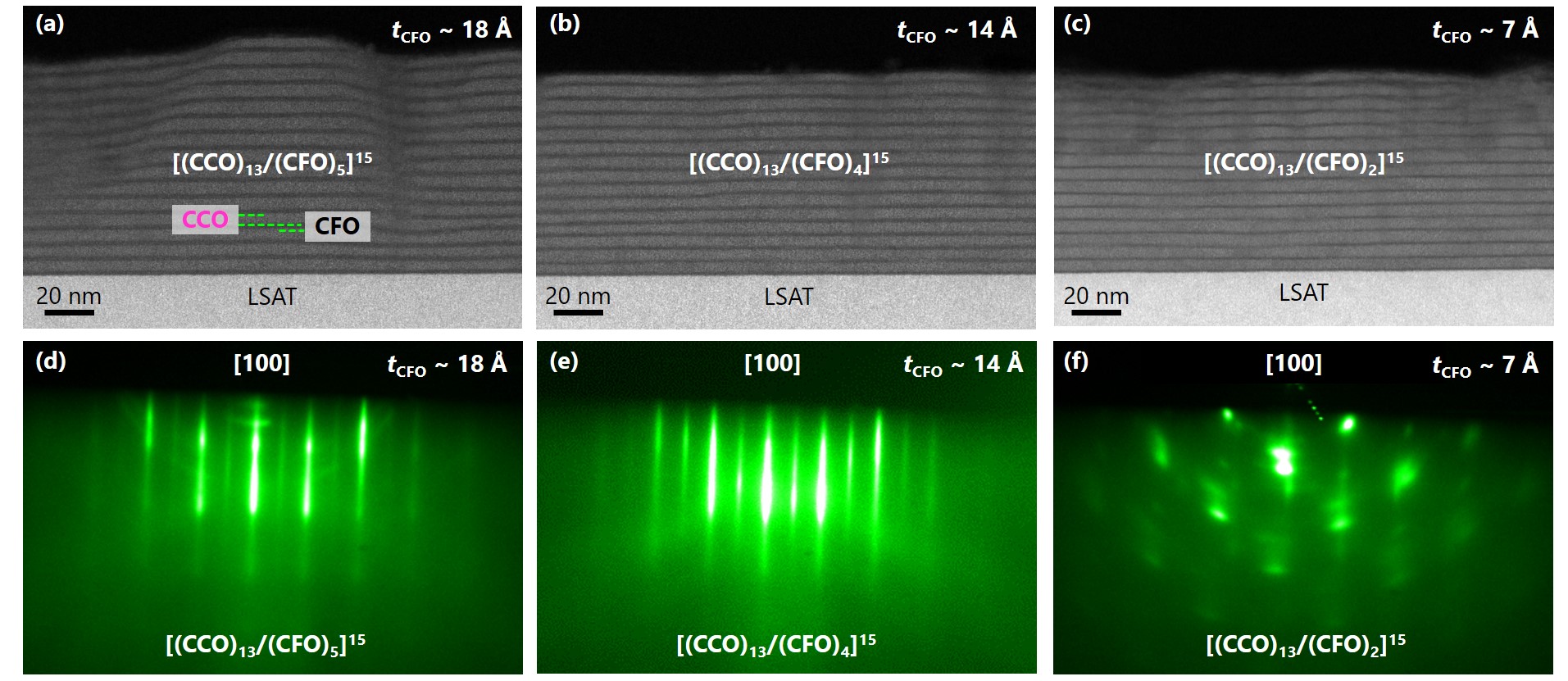}%
\caption{Low-magnification HADDF-STEM and RHEED images of [(IL-CaCuO$_{2}$)$_{13}$/(Ca$_{2}$Fe$_{2}$O$_{5}$)$_{m}$]$^{15}$ superlattices with different $m$. $t_{\text{CFO}}$ is 18, 14, 7\,\AA{ }for (a) and (d), (b) and (e), and (c) and (f), respectively. Note, that the apparent thickness of the IL-CaCuO$_{2}$ layers appears larger in (c) than in (b). Accordingly, the total thicknesses of the superlattices are nearly identical.\label{fig2}}
\end{figure*}

\section{RESULTS AND DISCUSSION}
In general, coherent growth of two or more materials is desirable for the synthesis of superlattices. Besides, this is also important for the stabilization of the IL-phase in superlattice form because IL-cuprates are barely stabilized by epitaxy. Figures \ref{fig2}(a-c) show low-magnification HAADF-STEM images of [(IL-CaCuO$_{2}$)$_{n}$/(Ca$_{2}$Fe$_{2}$O$_{5}$)$_{m}$]$^{N}$ superlattices  with different $m$ ($t_{\text{CFO}}$) to visualize alternate stacking of IL-CaCuO$_{2}$ and Ca$_{2}$Fe$_{2}$O$_{5}$ layers along the growth direction. The Ca$_{2}$Fe$_{2}$O$_{5}$ layers appear darker than the IL-CaCuO$_{2}$ layers in Fig.\ref{fig2}(a-c). This may seem odd because the HAADF intensity of the atomic columns is proportional to the atomic number $Z$ (here, $Z$ = 29 for Cu $>$ 26 for Fe $>$ 20 for Ca). One explanation for this observation is that appropriate defocusing conditions could be different between them. The superlattices with $t_{\text{CFO}}$ $\leq$ 14\,\AA{ }show higher thickness uniformity than those with $t_{\text{CFO}}$ $\sim$ 18\,\AA. This thickness dependency may be relevant to the critical thickness of Ca$_{2}$Fe$_{2}$O$_{5}$. In the absence of strain, Ca$_{2}$Fe$_{2}$O$_{5}$ crystallizes in an orthorhombic space group ($Pnma$), whereas IL-CaCuO$_{2}$ is tetragonal. Therefore, fully strained-Ca$_{2}$Fe$_{2}$O$_{5}$ grown on IL-CaCuO$_{2}$ is expected to be tetragonal when $t_{\text{CFO}}$ is thin enough. Based on our experiments, we can expect that coherent growth of Ca$_{2}$Fe$_{2}$O$_{5}$ in the superlattices can be allowed for $t_{\text{c}}$ $\leq$ 14\,\AA{ }($m$ $\leq$ 4). For $m$ $>$ 4, Ca$_{2}$Fe$_{2}$O$_{5}$ would be grown in a relaxed manner and this will lead to distortions in the subsequent IL-CaCuO$_{2}$ layers. 

\begin{figure}
\includegraphics[width=8.6cm]{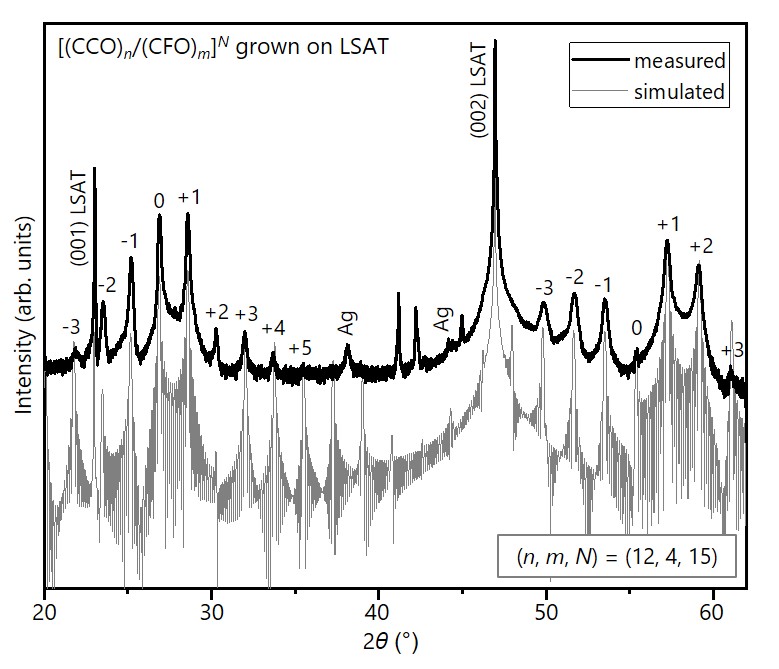}%
\caption{XRD pattern of [(IL-CaCuO$_{2}$)$_{n}$/(Ca$_{2}$Fe$_{2}$O$_{5}$)$_{m}$]$^{N}$ superlattices with ($n$, $m$, $N$) = (13, 4, 15) grown on (001)LSAT substrate. For comparison, simulated XRD pattern is also shown in grey. ($n$, $m$, $N$) = (12, 4, 15) gives a best fit to the measured XRD pattern. \label{fig3}}
\end{figure}

On the other hand, the STEM image for $t_{\text{CFO}}$ $\sim$ 7\,\AA{}[Fig.\ref{fig2}(c)] shows that the superlattices gradually degrade with increasing $N$. $t_{\text{CFO}}$ of each Ca$_{2}$Fe$_{2}$O$_{5}$ layers varied by more than 50$\%$ (Fig. S1 in the Supplemental Material), resulting in severe deterioration in subsequent IL-CaCuO$_{2}$ layers. Figures \ref{fig2}(d-f) show a comparison of RHEED images of IL-CaCuO$_{2}$ top layer between the superlattices with different $t_{\text{CFO}}$. A ringy and spotty pattern observed for $t_{\text{CFO}}$ $\sim$ 7\,\AA{ }most likely originates from the co-formation of thermodynamically stable Ca$_{2}$CuO$_{3}$ instead of IL-CaCuO$_{2}$. Again, this observation highlights the importance of epitaxy to stabilize the IL phase. The $t_{\text{CFO}}$ inhomogeneity observed for $t_{\text{CFO}}$ $\sim$ 7\,\AA{ }may be associated with the initial growth mode of Ca$_{2}$Fe$_{2}$O$_{5}$ on IL-CaCuO$_{2}$ but this is outside the scope of this paper. Instead, the further structural investigation will be done for the $t_{\text{CFO}}$ $\sim$ 14\,\AA .

The cross-sectional HAADF-STEM image for $t_{\text{CFO}}$ $\sim$ 14\,\AA{ }shows homogeneous stacking of the $N$ = 15 supercells, at least over the 200 nm range shown in Fig.\ref{fig2}(b). The surface structure of the superlattices was assessed by RHEED, as shown in Fig.\ref{fig2}(e). Clear streaks were recorded throughout the entire growth of the superlattices, a feature of layer-by-layer growth modes. The thickness of the supercell ($\Lambda$) was estimated to be 5.6$\pm$0.2\,nm from two-dimensional fast Fourier transformation of the real-space lattice data shown in Fig.\ref{fig2}(b). This value well coincides with the $\Lambda$ calculated from the diffraction angles of the satellite peaks observed in 2$\theta$/$\omega$ scan (Fig.\ref{fig3}). Based on the XRD pattern, the $\Lambda$ is found to be 5.46$\pm$0.10\,nm. We carried out computer simulations of XRD patterns using the software program InteractiveXRDFit \cite{Lichtensteiger:vh5084}. The simulated XRD pattern is also shown in Fig.\ref{fig3}. This program does not take influences of surface and interface roughness, thickness fluctuation, and defects on the diffraction peaks into account. Therefore, a broadening of the diffraction peaks and damping of higher-order satellites seen in the experimental data are not reproduced in the simulation results. Nonetheless calculated positions of the main and satellite peaks and their relative intensity coincide well with the experimentally determined diffraction data. The layer thicknesses of $t_{\text{CCO}}$ = 38.3 ($n$ $\sim$ 12) and $t_{\text{CFO}}$ = 14.8\,\AA{ }($m$ $\sim$ 4) were estimated from XRD simulation. These values are close to the ($n$, $m$) of (13, 4) obtained from HAADF-STEM images. Since XRD is a measurement technique sensitive to the coherent volume of the sample, the local structures captured by STEM can be considered to be representative over the whole range of the sample dimensions ($\sim$ 6\,mm$\times$3\,mm).

\begin{figure}
\includegraphics[width=8.6cm]{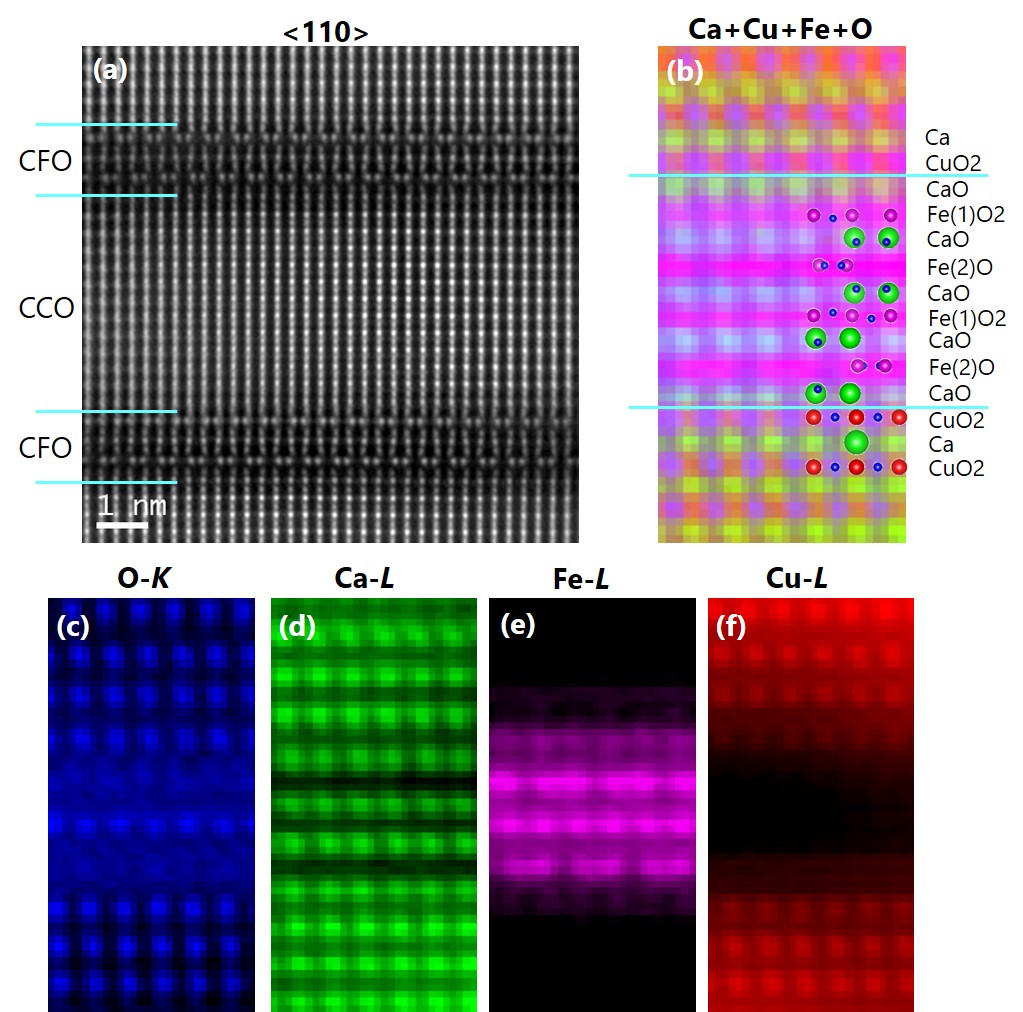}%
\caption{Atomically-resolved HAADF-STEM image and elementally-resolved EELS maps of [(IL-CaCuO$_{2}$)$_{n}$/(Ca$_{2}$Fe$_{2}$O$_{5}$)$_{m}$]$^{N}$ superlattices with ($n$, $m$, $N$) = (13, 4, 15). (a) Cross-sectional HAADF-STEM image along the [110] direction of IL-CaCuO$_{2}$. (b) Overlay of respective EELS elemental maps on (c) O $K$, (d) Ca $L$, (e) Fe $L$, (f) Cu $L$ edges. In (b), the color of each atom corresponds to one used in (c-f).\label{fig4}}
\end{figure}

In order to obtain structural information at an atomic level, we performed high-resolution HAADF-STEM combined with EELS on [(IL-CaCuO$_{2}$)$_{n}$/(Ca$_{2}$Fe$_{2}$O$_{5}$)$_{m}$]$^{N}$ superlattices with ($n$, $m$, $N$) = (13, 4, 15). Fig.\ref{fig4}(a) shows atomically resolved HAADF-STEM image. The observed atomic columns correspond to Ca and Cu of the IL and Ca and Fe of brownmillerite structures. The in-plane epitaxial relationship between IL-CaCuO$_{2}$ and Ca$_{2}$Fe$_{2}$O$_{5}$ is IL-CaCuO$_{2}$[100]/Ca$_{2}$Fe$_{2}$O$_{5}$[100]$_{\text{p}}$ [see Fig.\ref{fig1}(a,b)]. In the HAADF-STEM image, oxygen is barely visible. To visualize the position of oxygen, EELS elemental maps were recorded and plotted as shown in Fig.\ref{fig4}(b). The observed oxygen ordering coincides with oxygen positions in the IL and brownmillerite structures [Fig.\ref{fig4}(c)]. Owing to the orthorhombic symmetry of the brownmillerite unit cell, the distances between neighboring Ca atoms are not equal along the growth direction [Fig.\ref{fig4}(d)]. This reflects the brownmillerite structure stemming from alternate stacking of octahedral Fe(1)O$_{6}$ and tetrahedral Fe(2)O$_{4}$ as shown in Fig.\ref{fig1}(b). As shown in Fig.\ref{fig4}(e,f), there is no occurrence of an Cu/Fe interdiffusion at the interfaces. We can conclude that the interfaces between IL-CaCuO$_{2}$ and Ca$_{2}$Fe$_{2}$O$_{5}$ layers are atomically abrupt. The elementally resolved EELS maps are a direct proof that synthesized superlattices are actually made out of IL-CaCuO$_{2}$ and brownmillerite Ca$_{2}$Fe$_{2}$O$_{5}$.

\begin{figure}
\includegraphics[width=8.6cm]{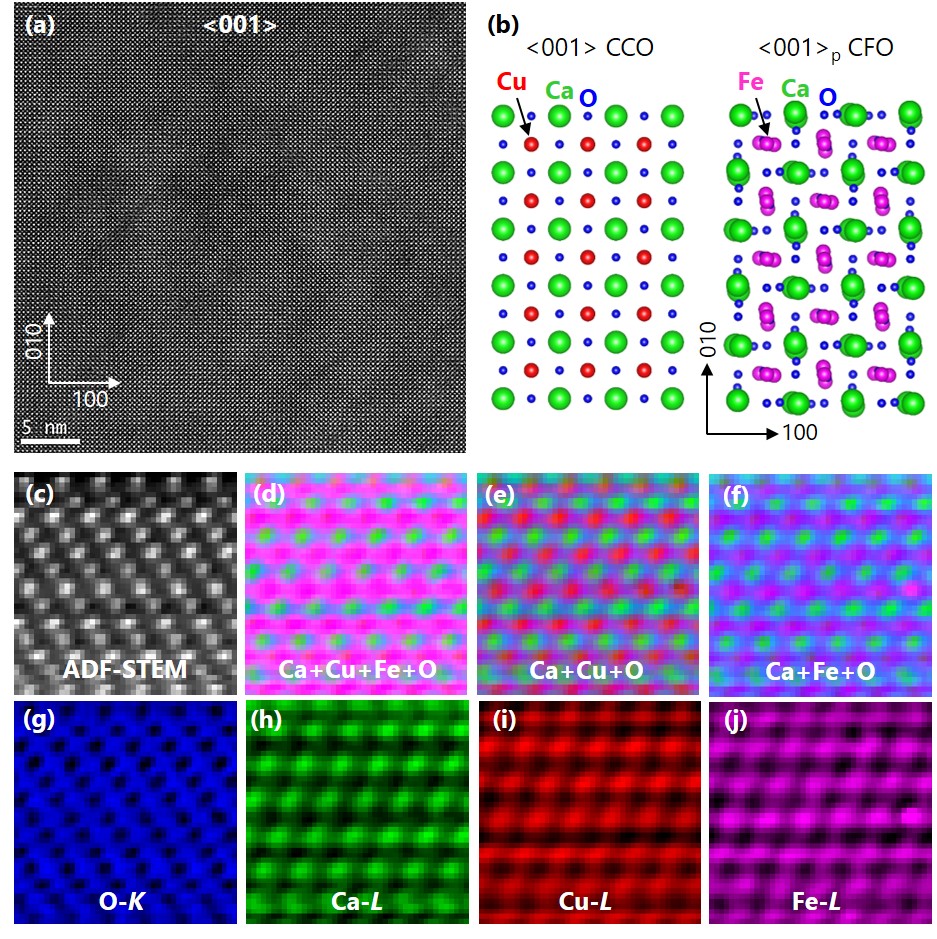}%
\caption{In-plane STEM-EELS on [(IL-CaCuO$_{2}$)$_{n}$/(Ca$_{2}$Fe$_{2}$O$_{5}$)$_{m}$]$^{N}$ superlatties with ($n$, $m$, $N$) = (13, 4, 15). (a) HAADF-STEM image and (b) schematic atomic arrangement of IL-CaCuO$_{2}$ as well as Ca$_{2}$Fe$_{2}$O$_{5}$ along the [001] direction. (c) Annular dark field image and (d-j) corresponding elementally-resolved EELS maps along the [001] direction. Overlay of EELS maps on O $K$, Ca $L$, Cu $L$, and Fe $L$ (d), O $K$, Ca $L$, and Cu $L$ (e), O $K$, Ca $L$, and Fe $L$ edges (f). Individual EELS maps on O $K$ (g), Ca $L$ (h), Cu $L$ (i), and Fe $L$ edges (j), respectively. \label{fig5}}
\end{figure}

Interstitial oxygen defects are commonly seen in oxide materials, and IL and brownmillerite oxides are no exception. In the case of IL-cuprates, the apex position of copper adjacent to Ca/Sr ions can be a possible interstitial site. Neutron powder diffraction measurement on IL-Sr$_{0.9}$La$_{0.1}$CuO$_{2}$ pointed out the absence of such interstitial oxygen ions \cite{PhysRevB.47.14654} whereas several works on thin film imply their existence \cite{KARIMOTO2002127, ADACHI199214, doi:10.1063/1.1410872, doi:10.1063/1.2339840}. A recent report by D. Castro $et$ $al.$ claims that the interstitial site at the interface between IL-CaCuO$_{2}$ and SrTiO$_{3}$ is occupied by excess oxygen ions \cite{PhysRevLett.115.147001}. However, it seems safe to say that such defects are absent in the [(IL-CaCuO$_{2}$)$_{n}$/(Ca$_{2}$Fe$_{2}$O$_{5}$)$_{m}$]$^{N}$ superlattices shown here, as one can see in Fig.\ref{fig4}(c).

In contrast to IL-cuprates, brownmillerite ferites are known to incorporate oxygen ions and used as a starting material for topotactic phase transformations into the perovskite phase \cite{TAKEDA197861, MORIMOTO199766, PhysRevB.62.844, PhysRevLett.91.156405, PhysRevMaterials.2.015002, doi:10.1063/1.1445805, doi:10.1063/1.5002672}. Since the perovskite phase has an unusual high valence state Fe$^{4+}$, this phase transformation is done under very strong oxidizing conditions, $e.g.$ with the use of oxygen sources such as KClO$_{4}$ at high-pressure (2-3\,GPa), or oxygen plasma or ozone as an oxidizing agent. The oxidizing atmosphere during cooling, used in this study, may be similar to that used for the synthesis of perovskite CaFeO$_{3}$ thin films \cite{PhysRevMaterials.2.015002}. However, the possibility that brownmillerite Ca$_{2}$Fe$_{2}$O$_{5}$ layers in our superlattices are fully transformed into the perovskite phase can be easily ruled out, as shown in the elementally resolved STEM-EELS maps [Fig.\ref{fig4}]. On the other hand, excess oxygen ions can be accommodated in the brownmillerite Ca$_{2}$Fe$_{2}$O$_{5+\delta}$. Given the limited data for the oxygen nonstoichiometry ($\delta$) of brownmillerite Ca$_{2}$Fe$_{2}$O$_{5+\delta}$, $\delta$ has been reported to be close to zero or at most 0.08 \cite{shaula2013}. For 0.09 $\leq$ $\delta$ $\leq$ 0.8, the crystal structure is neither brownmillerite nor perovskite \cite{JM9950501909}. The Fe-$L$ EELS spectra taken from the Ca$_{2}$Fe$_{2}$O$_{5}$ layer in our superlattices shows a single peak similar to one reported for brownmillerite Ca$_{2}$Fe$_{2}$O$_{5}$\cite{doi:10.1063/1.3610526, Jo2020} (Fig. S2 in the supplemental Material), unlike two-peak structure observed for oxygen nonstoichiometric phases \cite{Galakhov2010}. These results suggest that the $\delta$ in the superlattices is expected to be negligible even after cooling in atomic oxygen.

Unlike anionic defects, cationic defect formation can be easily traced by STEM imaging. Direct comparison of structural imaging data between [(IL-CaCuO$_{2}$)$_{n}$/(Ca$_{2}$Fe$_{2}$O$_{5}$)$_{m}$]$^{N}$ superlattices and bare IL-CaCuO$_{2}$ thin films allows us to conclude that the crystal quality of IL-CaCuO$_{2}$ has significantly improved by inserting the Ca$_{2}$Fe$_{2}$O$_{5}$ layers as follows. The creations of large scale (tens of nm) V-shaped defects observed in bare IL-CaCuO$_{2}$ thin films \cite{JAP.124.2018.073905} are barely visible in [(IL-CaCuO$_{2}$)$_{n}$/(Ca$_{2}$Fe$_{2}$O$_{5}$)$_{m}$]$^{N}$ superlattices, as shown in Fig.\ref{fig2}.  In addition to cross-sectional observation, further investigation was done by in-plane STEM imaging. Figure \ref{fig5} shows in-plane HAADF-STEM and elementally resolved EELS maps of a [(IL-CaCuO$_{2}$)$_{n}$/(Ca$_{2}$Fe$_{2}$O$_{5}$)$_{m}$]$^{N}$ superlattice along the (001) direction. The observed HAADF-STEM image does not show any in-plane and out-of-plane line defects, in stark contrast to bare IL-CaCuO$_{2}$ thin films where (100) and (010) anti-site boundary defects are observed \cite{Krockenberger2021}. Such defects are most likely triggered by charge imbalances caused by local defects \cite{JAP.124.2018.073905, PhysRevMaterials.3.064803, Krockenberger2021}. The insertion of Ca$_{2}$Fe$_{2}$O$_{5}$ layers increases the degree of cation ordering in IL-CaCuO$_{2}$. Elementally resolved in-plane EELS maps, shown in Fig.\ref{fig5} corroborate the scenario that IL-CaCuO$_{2}$ sandwiched by the Ca$_{2}$Fe$_{2}$O$_{5}$ layers is nearly monolithic.  This suggests that inserted layers may play a role in the CRLs to compensate the charge imbalances triggered by point defect formation during growth of
 the [(IL-CaCuO$_{2}$)$_{n}$/(Ca$_{2}$Fe$_{2}$O$_{5}$)$_{m}$]$^{N}$ superlattices without causing the distortion in IL-CaCuO$_{2}$. 

\begin{figure}
\includegraphics[width=8.6cm]{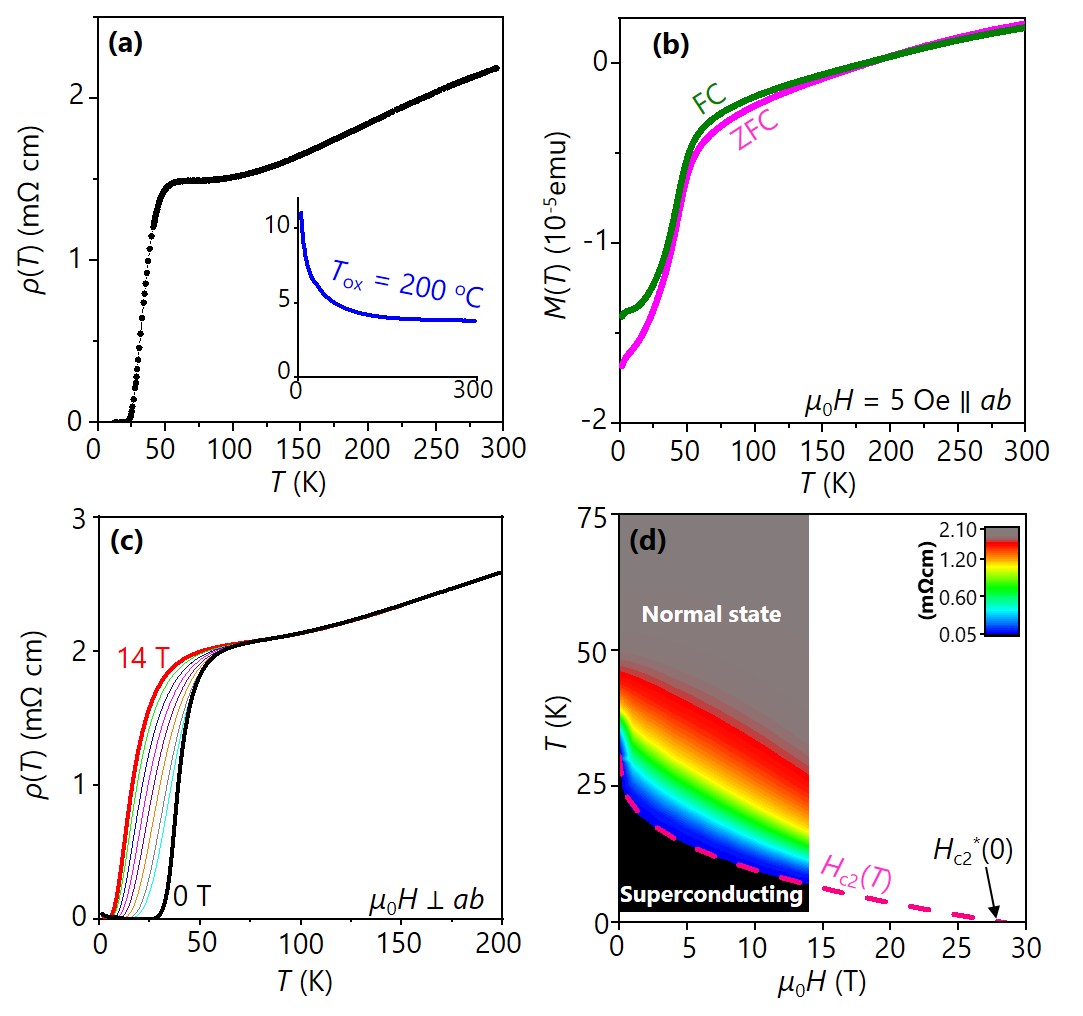}%
\caption{Superconducting properties of [(IL-CaCuO$_{2}$)$_{n}$/(Ca$_{2}$Fe$_{2}$O$_{5}$)$_{m}$]$^{N}$ superlattices. (a) Temperature dependences of resisvitiy ($\rho$) for [(IL-CaCuO$_{2}$)$_{n}$/(Ca$_{2}$Fe$_{2}$O$_{5}$)$_{m}$]$^{N}$ superlattices with ($n$, $m$, $N$) = (17, 4, 11) cooled in atomic oxygen down to $T_{\text{ox}}$ = 140\textdegree C. For comparison, the $\rho$-$T$ curve for [(IL-CaCuO$_{2}$)$_{13}$/(Ca$_{2}$Fe$_{2}$O$_{5}$)$_{3}$]$^{11}$ superlattices with $T_{\text{ox}}$ = 200\textdegree C is shown in the inset. (b) Field cooled (FC) and zero field cooled (ZFC) magnetization as a function of temperature for [(IL-CaCuO$_{2}$)$_{n}$/(Ca$_{2}$Fe$_{2}$O$_{5}$)$_{m}$]$^{N}$ superlattices with ($n$, $m$, $N$) = (17, 4, 11). The external field ($\mu_{0}H$) of 5\,Oe was applied parallel to the film surface. (c) Temperature dependeces of $\rho$ for [(IL-CaCuO$_{2}$)$_{n}$/(Ca$_{2}$Fe$_{2}$O$_{5}$)$_{m}$]$^{N}$ superlattices with ($n$, $m$, $N$) = (17, 4, 6) under a magnetic field of 0, 1, 2, 4, 6, 8, 10, 12, and 14\,T applied perpendicular to the film surface. (d) Resistivity contour map on [(IL-CaCuO$_{2}$)$_{n}$/(Ca$_{2}$Fe$_{2}$O$_{5}$)$_{m}$]$^{N}$ superlattices with ($n$, $m$, $N$) = (17, 4, 6) as a fucntion of $\mu_{0}H$ and $T$. The dashed line is a fitting curve derived from $H_{\text{c2}}$($T$) = $H_{\text{c2}}^{*}$(0)(1-$\frac{T}{T_{\text{c}}}$)$^{\alpha}$ with $H_{\text{c2}}^{*}$(0) = 28.03$\pm$0.47\,T and $\alpha$ = 2.76$\pm$0.05. \label{fig6}}
\end{figure}

Any distortions in IL-CaCuO$_{2}$ would directly influence its superconductivity because of the crystallographic simplicity of the IL structure. IL-CaCuO$_{2}$ is known to exhibit an insulating transport behavior and reported resistivity value ranges from 70 m$\Omega$cm to 4 $\Omega$cm \cite{JAP112.2012, ThinSolidFilms.2005301, JAP.124.2018.073905}. However, the superlattices of [(IL-CaCuO$_{2}$)$_{n}$/(Ca$_{2}$Fe$_{2}$O$_{5}$)$_{m}$]$^{N}$ show metallic transport behavior followed by superconductivity below 50\,K [Fig.\ref{fig6}(a)]. The superconducting transition is also confirmed by magnetization ($M$) measurements, as shown in Fig.\ref{fig6}(b). Here, an external magnetic field was applied parallel to the film surface and therefore demagnetization coefficient is nearly zero. A clear diamagnetic shielding is observed from the $M$-$T$ curves implying robust superconductivity in the [(IL-CaCuO$_{2}$)$_{n}$/(Ca$_{2}$Fe$_{2}$O$_{5}$)$_{m}$]$^{N}$ superlattices. 
Figure \ref{fig6}(c) shows the resistivity as a function of temperature under magnetic fields up to $\mu_{0}H$ = 14\,T applied perpendicular to the film surface. $T_{\text{c}}$ gradually decreases with increasing $\mu_{0}H$ but superconductivity persists up to 14\,T. In Fig.\ref{fig6}(d), $B$-$T$ phase diagram of the [(IL-CaCuO$_{2}$)$_{n}$/(Ca$_{2}$Fe$_{2}$O$_{5}$)$_{m}$]$^{N}$ superlattices is shown using a contour map generated from the resistivity values. Here, we define the superconducting state to be $\rho$ values $<$ 0.05 m$\Omega$cm --- shown in Fig.\ref{fig6}(d) as filled black --- to minimize uncertainty due to the broad resistive transition and avoid any influence of the contact resistance of indium. One can see a pronounced positive curvature ($\frac{d^2}{dT^2}H_{\text{c2}}$($T$) $>$ 0) in the temperature dependence of the upper critical field $H_{\text{c2}}$($T$) near $T_{\text{c}}$. This behavior is inconsistent with a linear temperature dependence given by the single-band Ginzburg-Landau theory. Using  $H_{\text{c2}}$($T$) = $H_{\text{c2}}^{*}$(0)(1-$\frac{T}{T_{\text{c}}}$)$^{\alpha}$ where $H_{\text{c2}}^{*}$(0) is the upper limit of $H_{\text{c2}}$(0), we found $H_{\text{c2}}^{*}$(0) = 28.03$\pm$0.47\,T and $\alpha$ = 2.76$\pm$0.05 as best fitting parameters. The power law dependence and/or positive curvature on $H_{\text{c2}}$($T$) in the vicinity of $T_{\text{c}}$ have often been reported for high-$T_{\text{c}}$ cuprates \cite{PhysRevLett.64.599, PhysRevB.46.5581, PhysRevB.54.1251} and other superconducting materials such as MgB$_{2}$, Ba$_{1-x}$K$_{x}$BiO$_{3}$ \cite{Askerzade_2002, PhysRevB.54.6133, PhysRevB.49.3502}. For high-$T_{\text{c}}$ cuprates, such behaviors are mainly related to flux-creep dissipation or fluctuation effects. They are partly associated with layered structures of this system and the presence of defects and/or inhomogeneities, to some extent, for complex oxides such as cuprates. We note that the positive curvature in $H_{\text{c2}}$($T$) presented here is triggered by defects/phase inhomogeneities cannot be excluded. For comparison, the $H_{\text{c2}}$ measured for the [(IL-CaCuO$_{2}$)$_{n}$/(Ca$_{2}$Fe$_{2}$O$_{5}$)$_{m}$]$^{N}$ superlattices is larger than that of 13.9\,T reported for electron-doped IL-cuprates, where $T_{\text{c}}$ is higher rather than the superlattices presented here \cite{PhysRevB.66.214509}. Using $H_{\text{c2}}^{*}$(0) = 28.03\,T, we calculated a coherence length $\xi$ = ($\phi_{0}$/2$\pi H_{\text{c2}}^{*}(0)$)$^{1/2}$ for the [(IL-CaCuO$_{2}$)$_{n}$/(Ca$_{2}$Fe$_{2}$O$_{5}$)$_{m}$]$^{N}$ superlattices and obtained $\xi$ = 34.3\,\AA . In the in-plane HAADF-STEM image shown in Fig.\ref{fig5}(a), the presence of defects is barely seen at least over the scanned area (380\,\AA $\times$380\,\AA), implying that the [(IL-CaCuO$_{2}$)$_{n}$/(Ca$_{2}$Fe$_{2}$O$_{5}$)$_{m}$]$^{N}$ superlattices shown here might be in the clean limit.

The superconducting transition observed here is rather broad. In the $M$-$T$ curves, superconducting transition starts slightly higher than 50\,K, which approximately coincides with the onset temperature of 50\,K in the $\rho$-$T$ curve. Zero resistivity appears below 23\,K. The broad superconducting transition of about 27\,K might be influenced by how the superlattice was cooled after its synthesis. The superconducting superlattices shown here were obtained through a cooling process in atomic oxygen after the synthesis. Similar findings have been reported for IL-CaCuO$_{2}$ and SrTiO$_{3}$ superlattices \cite{PhysRevB.86.134524}. The [(IL-CaCuO$_{2}$)$_{n}$/(Ca$_{2}$Fe$_{2}$O$_{5}$)$_{m}$]$^{N}$ superlattices cooled under vacuum ($\sim$10$^{-8}$\,Torr) show insulating behavior and the resistivity value at 295\,K as high as several $\Omega$cm. The $\rho$ values decreases with $T_{\text{ox}}$ at which the irradiation of atomic oxygen stops during the cooling process [see the inset of Fig.\ref{fig6}(a)]. Superconductivity was observed in the superlattices for $T_{\text{ox}}$ $\leq$140\textdegree C. The superlattices shown here are cooled by switching off the substrate heater after synthesis and cooling from the synthesis temperature $T_{\text{s}}$ = 580-590\textdegree C to 300\textdegree C typically takes about 20 minutes and subsequent further cooling down to 140\textdegree C takes another 40 minutes. This long period of irradiation with atomic oxygen may cause damage to the film surface and/or inhomogeneous oxidation. Cross-sectional STEM images showing the top IL-CaCuO$_{2}$ layer of the superlattices support the former scenario, as shown in Fig. \ref{fig2}(a-c). One can expect that further optimization of the oxidation process after growth leads to a sharp superconducting transition, and moreover it may boost $T_{\text{c}}$.

As shown above, strong oxidation down to $T_{\text{ox}}$ $\leq$ 140\,$^\circ\mathrm{C}$ is required for the induction of superconductivity in the [(IL-CaCuO$_{2}$)$_{n}$/(Ca$_{2}$Fe$_{2}$O$_{5}$)$_{m}$]$^{N}$ superlattices.  Several possible scenarios explaining our finding might involve carrier doping via incorporation of excess oxygen ions at the interstitial sites of either IL-CaCuO$_{2}$, as is the case with IL-CaCuO$_{2}$/SrTiO$_{3}$ superlattices \cite{PhysRevLett.115.147001}, or Ca$_{2}$Fe$_{2}$O$_{5}$. However, high-resolution STEM-EELS analysis shown in Fig.\ref{fig4} clearly rules out the incorporation of excess oxygen ions in the IL-CaCuO$_{2}$ layers. Assuming that $\delta$ $\ll$ 0.08 in Ca$_{2}$Fe$_{2}$O$_{5+\delta}$ can produce 2$\delta$ holes in the CuO$_{2}$ planes of IL-CaCuO$_{2}$, the carrier density of (IL-CaCuO$_{2}$)$_{13}$/(Ca$_{2}$Fe$_{2}$O$_{5+\delta}$)$_{4}$ superlattices is estimated to be $\leq$ 2$\times$10$^{20}$\,cm$^{-3}$. This value is at least one order of magnitude smaller than that typically found in doped multilayer cuprates \cite{10.1007/978-4-431-68195-3_54, YASUSHIIDEMOTO1995123}. Taking into account bulk-like superconducting shielding observed in [(IL-CaCuO$_{2}$)$_{n}$/(Ca$_{2}$Fe$_{2}$O$_{5}$)$_{m}$]$^{N}$ superlattices [Fig.\ref{fig6}(b)], it is unlikely that such small $\delta$ for large $n$ $\geq$ 13 is responsible for the induction of superconductivity. Instead, the presence of oxygen vacancies on the CuO$_{2}$ planes in IL cuprates have been suggested for electron-doped IL-cuprates by R$^{3+}$-substitution \cite{PhysRevB.92.035149} and elimination of such defects by the irradiation of atomic oxygen down to $T_{\text{ox}}$ is necessary for the induction of superconductivity \cite{doi:10.1063/1.2339840}. Although high-resolution $in{ }situ$ scanning tunneling microscopy study pointed out a buckling of CuO$_{2}$ planes in non-superconducting IL-SrCuO$_{2}$ rather than oxygen vacancies, it was shown that post-annealing under ultra-high vacuum results in a removal of oxygen ions and formation of stripe defects \cite{PhysRevB.97.245420} which is distinct from cationic stripe formation found in IL-CaCuO$_{2}$, Ca$_{1-x}$R$_{x}$CuO$_{2}$, and Ca$_{1-x}$Sr$_{x}$CuO$_{2}$ \cite{Krockenberger2021}. Such defects are hard to detect even with high-resolution STEM technique because they usually exist as random point defects. We suggest that the oxidation throughout cooling process prevents oxygen vacancies in CuO$_{2}$ planes and is prerequisite for high-$T_{\text{c}}$ superconductivity in IL-CaCuO$_{2}$.

\begin{figure}
\includegraphics[width=8.6cm]{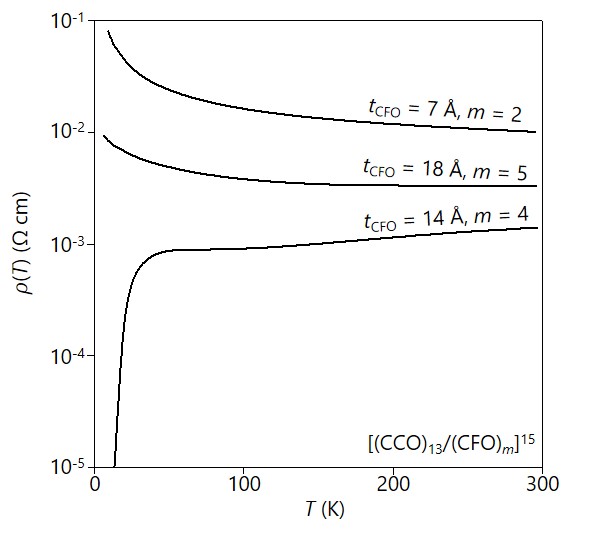}%
\caption{Temperature dependent resistivity of [(IL-CaCuO$_{2}$)$_{13}$/(Ca$_{2}$Fe$_{2}$O$_{5}$)$_{m}$]$^{15}$ superlattices with $t_{\text{CFO}}$ = 7, 14, and 18\,\AA ($m$ = 2, 4, and 5). $T_{\text{ox}}$ is common and fixed at 140\textdegree C. \label{fig7}}
\end{figure}

Superconductivity is observed for 11 $\leq$ $t_{\text{CFO}}$ $\leq$ 14\,\AA , corresponding to 3 $\leq$ $m$ $\leq$ 4. However, the [(IL-CaCuO$_{2}$)$_{n}$/(Ca$_{2}$Fe$_{2}$O$_{5}$)$_{m}$]$^{N}$ superlattices with either $m$ $<$ 3 or $m$ $>$ 4 remain insulating even after cooling down to $T_{\text{ox}}$ = 140\textdegree C under flowing atomic oxygen although their resistivity value is at least two orders of magnitude smaller than that for bare IL-CaCuO$_{2}$ thin films (Fig.\ref{fig7}). This high sensitivity to $m$ again argues against the notion that the excess oxygen is introduced in the Ca$_{2}$Fe$_{2}$O$_{5+\delta}$ layers and provides hole carriers in the CuO$_{2}$ planes of IL-CaCuO$_{2}$. Our results imply that superconductivity in IL-CaCuO$_{2}$ is very sensitive to structural distortions, probably owing to the lack of CRLs. Since Ca$_{2}$Fe$_{2}$O$_{5}$ turned out to serve as CRLs, the approach of IL-CaCuO$_{2}$-based superlattice formation is promising towards the synthesis of higher-$T_{\text{c}}$ superconductors.

\section{CONCLUSION}
We have shown the synthesis of superconducting artificial superlattices of IL-CaCuO$_{2}$ and brownmillerite Ca$_{2}$Fe$_{2}$O$_{5}$ using MBE. In superlattices, the IL-CaCuO$_{2}$ is a superconducting layer and the Ca$_{2}$Fe$_{2}$O$_{5}$ responses as a CRL so as to tolerate charge imbalances introduced during growth. Superlattices have a significant effect on stabilizing IL-CaCuO$_{2}$ without the creation of cationic defects disrupting the CuO$_{2}$ planes. In addition, the oxygen-stoichiometry in the CuO$_{2}$ planes governs the emergence of superconductivity and the prevention of oxygen vacancy formation in CuO$_{2}$ planes will be the common challenge for the synthesis of superconducting superlattices using different material combinations, and more importantly for the enhancement of $T_{\text{c}}$.

\begin{acknowledgments}
\section{ACKNOWLEDGEMENT}
The authors gratefully acknowledge Michio Naito for discussion. We thank Takayuki Ikeda for his contribution in STEM and EELS measurements. 
\end{acknowledgments}

\bibliographystyle{apsrev4-1}
\bibliography{references}

\end{document}